\documentclass[aps,twocolumn,showpacs]{revtex4}
\usepackage{dcolumn}
\usepackage{graphicx}
\usepackage{amsmath}
\usepackage{amsfonts}
\usepackage{amssymb}
\usepackage{psfrag}
\usepackage{wrapfig}
\usepackage{subfigure}
\usepackage{makeidx}
\usepackage{bm}
\usepackage{epsf}
\DeclareMathOperator{\sech}{sech}

\begin{document}

\title{The Mechanism of Kuznetsov-Ma Breather}
\author{Li-Chen Zhao$^{1,2}$}\email{zhaolichen3@nwu.edu.cn}
\author{Liming Ling$^3$}
\author{Zhan-Ying Yang$^{1,2}$}

\address{$^{1}$School of Physics, Northwest University, Xi'an, 710069, China}
\address{$^{2}$Shaanxi Key Laboratory for Theoretical Physics Frontiers, Xi'an, 710069, China}
\address{$^{3}$School of Mathematics, South China University of Technology, Guangzhou 510640, China}

\date{\today}
\begin{abstract}
We discuss how to understand the dynamical process of Kuznetsov-Ma breather, based on some basic physical mechanisms.  It is shown that dynamical process of Kuznetsov-Ma breather involves at least two distinctive mechanisms:  modulational instability, and the interference effects between a bright soliton and a plane wave background. Our analysis indicates that modulational instability plays dominant roles in mechanism of Kuznetsov-Ma breather admitting weak perturbations, and the interference effect plays dominant role for the Kuznetsov-Ma breather admitting strong  perturbations. For intermediate cases, the two mechanisms are both involved greatly. These characters provide a possible way to understand the evolution of strong perturbations on a plane wave background.


\end{abstract}
\pacs{05.45.Yv, 02.30.Ik, 42.65.Tg}
\maketitle


\section{Introduction}
Recently, localized waves on plane wave background (PWB) became intense since their dynamics is related with freak wave, mainly including  Akhmediev breather (AB) \cite{AB}, Peregrine rogue wave (RW) \cite{RW},  and K-Mb \cite{K-M}. They have been excited in real nonlinear systems, such as optical lase field in fiber \cite{Kibler,Dudley,Kibler2}, and  water wave tank \cite{Chabchoub}. The underlying mechanism of their dynamics has been paid much attention after investigating their dynamics through deriving analytical solutions. The mechanism  is mainly refer to how to understand the dynamical process of  these localized waves in simple ways, based on some general physical properties. Modulational instability (MI), which is  associated with the growth of weak perturbations on a PWB, has been seen as mechanism of RWs and ABs \cite{Dudley}. Furthermore, baseband MI or MI with
resonant perturbations is found to play an essential role in RW excitations
\cite{Baronio1,Baronio3,zhaoling}, and many efforts have been paid to  the general nature of nonlinear stage of MI to understand MI more systematically
\cite{NMI,Zakharov}.  Furthermore, the underlying mechanisms for forming different spatial-temporal structure of fundamental RWs or ABs  were uncovered very recently \cite{Lingzhao}. All these studies further deepen our understanding on RW and AB dynamics greatly. However,  MI can not explain the dynamical process of   K-Mb well, partly because of MI usually fails to explain evolution of strong perturbations on a PWB \cite{biondini2}. Meanwhile, it should be noted that nonlinear MI has been proposed to predict and explain the evolution of strong localized perturbations with continuous inverse scattering spectrum  \cite{NMI,biondini3}. We mainly focus on how to understand the dynamical process of K-Mb, based on some general physical mechanisms.

K-Mb is generally  a nonlinear superposition of a bright soliton (BS) and a PWB, since the bright soliton related term  depends on plane wave term. Interestingly, it can be  written in a linear superposition form of them at some special moments. Therefore the BS term could be seen as a ``perturbation" term on the PWB. When BS's amplitude is much smaller than the background's, the K-Mb's dynamics can be explained well by MI  \cite{zhaoling}. In a limit case, the soliton's amplitude
tend to be zero, K-Mb will tend to a RW. These cases for weak perturbations are surely explained well by linear stability analysis on a plane wave background. However, the soliton's amplitude can be much larger than the background's for K-Mb, which makes the linear stability analysis usually do not hold anymore.  Moreover, we demonstrated that breathers could exist in modulational stability (MS) regime which could not be reduced to RW anymore \cite{defnls}, and anti-dark soliton was reported to exist in MI regime with some fourth-order effects \cite{AD}. These striking localized waves can not be explained by MI. We would like to explain the mechanism of K-Mb to provide a reasonable understanding of them, since these localized waves are all related with K-Mb excitation.

In this paper, we discuss how to understand the dynamical process of K-Mb, based on some general physical mechanisms. An approximation form is introduced to describe the interference effects, which is one of the fundamental properties in both classical wave and quantum theory. The analysis results suggest that the dynamical process of K-Mb involves at least two distinctive mechanisms, MI and interference effects. Since the two mechanisms can be used to explain the dynamical characters of K-Mb well, they are seen as the mechanism of K-Mb.  The interference effects are between a bright soliton and plane wave. For K-Mb admitting weak perturbation cases,  the oscillation period can be explained by nonlinear interference effects, and the amplitude oscillation can be understood well by MI. For K-Mb admitting strong perturbation, the oscillation period and amplitude oscillation can be both explained by linear interference effects. The weak and strong perturbations are clarified by the ratio $p/s<< 1$ and $p/s>>1$ respectively ($p$ and $s$ denote BS amplitude and PWB amplitude).  For intermediate cases, it is expected that the two mechanisms both are involved greatly. These understandings on K-Mb mechanism are summarized in Fig. 1. Additionally, the interference effects can be also used to explain the anti-dark soliton in MI regime and breather excitation in MS regime reported before \cite{defnls,AD}.

\section{Analysis on  Kuznetsov-Ma  breather}
For simplicity and without losing generality, we would like to begin with a generalized K-Mb solution which has been given widely for the well-known nonlinear Schr\"{o}dinger
equation  $i \Psi_{t}+\frac{1}{2}\Psi_{xx}+ |\Psi|^2 \Psi=0$. Similar discussion can be extended conveniently to K-Mb like breather in other nonlinear systems. Its explicit form can be  written as follows,
\begin{eqnarray}
\Psi&=&  \left[s-\frac{2(b^2-s^2)\cos(\xi t)+i \xi \sin(\xi t)}{b \cosh(2 x \sqrt{b^2-s^2})-s \cos(\xi t)}\right] e^{i s^2 t},
\end{eqnarray}
where $\xi=2 b \sqrt{b^2-s^2}$.
The parameter $b>=s$ determines the initial nonlinear localized wave's
shape, and $s$ is the background amplitude for
localized waves. It should be noted that the solution mainly has two terms.  The first term is a PWB, and the other term corresponds to a BS related term. Since the second term depends on the PWB generally, the K-Mb is a nonlinear superposition of BS and PWB. However, for $t=\frac{\pi+2 n \pi}{4 b \sqrt{b^2-s^2}}$ ($n$ is an integer), the K-Mb can be written as
 \begin{eqnarray}
\Psi&=& s e^{i \phi}-i 2\sqrt{b^2-s^2}\ \sech (2 \sqrt{b^2-s^2} x ) \ e^{i \phi },
\end{eqnarray}
where $\phi=s^2 \frac{\pi}{4 b \sqrt{b^2-s^2}}$. This can be seen as  a linear superposition of a PWB and a BS. The BS's amplitude is $2 \sqrt{b^2-s^2}$, which can be varied conveniently to investigate the evolution of strong perturbations on PWB. If the soliton amplitude is much smaller than the PWB amplitude, the BS term can be seen as the perturbation term $f_{pert}$ in the typical linear instability analysis on a PWB. This idea has been used to clarify the relations between MI and several nonlinear excitations \cite{zhaoling}. When the amplitude of soliton perturbation tend to be zero, the K-MB dynamical process tend to be a RW which admits rational amplification form \cite{Kibler,Dudley,Kibler2}.  Fourier analysis of the localized perturbation suggest that both RW and K-Mb admit dominant wave vector at resonant one with the background \cite{zhaoling,biondini2} (the wave vector is called according to the spatial coordinate $x$). The MI analysis predicts the perturbations with resonant wave vector admit rational amplification form, namely  $1 + i 2 s^2 t $.    Therefore, the amplification of K-Mb admitting weak soliton perturbations can be explained well by MI.   However, when the soliton's amplitude is comparable with the background's, the linear stability does not hold anymore. For example,  we demonstrated that breathers could exist in modulational stability (MS) regime which could not be reduced to RW anymore \cite{defnls}, and anti-dark soliton was reported to exist in MI regime with some fourth-order effects \cite{AD}. The anti-dark soliton's existence indicate that strong perturbations are possible to be stable in MI regime. The breather demonstrate striking oscillation characters in the MS regime. These localized waves' dynamical process can not be explained at all, based on MI mechanism. Therefore, we mainly try to find a possible way to understand these cases with strong perturbations in the following parts.

\begin{figure}[htb]
\centering
\label{fig:1}
{\includegraphics[height=65mm,width=78mm]{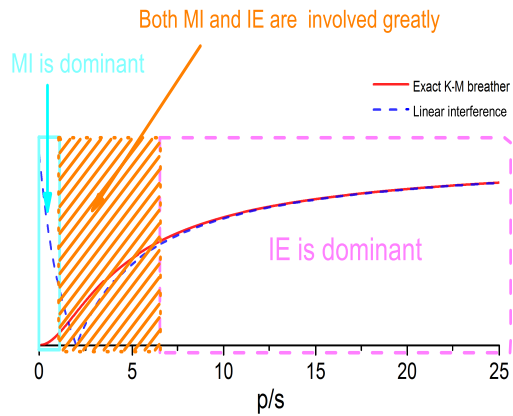}}
\caption{(color online) A qualitative description on the mechanisms of K-Mb. MI plays dominant role for K-Mb admitting weak perturbations, and the interference effects (IE) play dominant role for K-Mb admitting  strong perturbations. The weak and strong perturbations are clarified by the ratio $p/s<< 1$ and $p/s>>1$ respectively (p and s denote perturbation amplitude and background amplitude respectively). For intermediate cases ($p$ is comparable with $s$), it is expected that the two mechanisms  are both involved greatly. For weak perturbation case and intermediate cases, the breathing period can be explained well by nonlinear interference effects. }
\end{figure}

  \begin{figure}[htb]
\centering
\label{fig:2}
{\includegraphics[height=55mm,width=78mm]{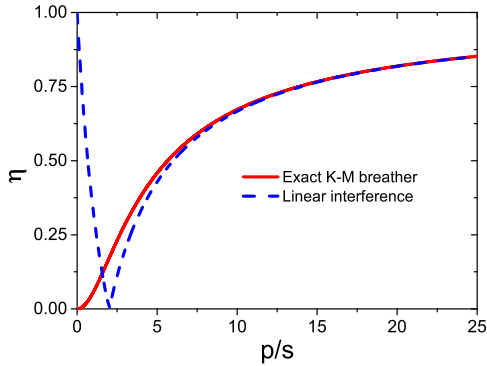}}
\caption{(color online) The oscillation amplitude characterization parameter $\eta$ vs $p/s$ (the ratio of perturbation soliton amplitude and background amplitude). The solid red line and blue dashed line correspond to the results of K-Mb and the linear interference case respectively. It is shown that the amplitude oscillation behavior of K-Mb does not agree with linear interference effects for weak soliton perturbations and the  intermediate cases, but the behavior agree perfectly with the prediction of linear interference effects for strong soliton perturbations. }
\end{figure}

We can see that even for the nonlinear superposition form, K-Mb tend to be a linear superposition form when the soliton amplitude is much larger than the background amplitude. This provides us a hint to understand the dynamics of soliton-type perturbations with large amplitudes on a plane wave background. We therefore introduce a linear superposition form of a BS with a generic form and a PWB according to the above linear form
 \begin{eqnarray}
\Psi'&=& s  e^{i s^2 t}- i p \  \sech(p x)\ e^{i  p^2 t/2},
\end{eqnarray}
 where $p$ is the soliton amplitude and it corresponds to the amplitude  $2 \sqrt{b^2-s^2}$ of soliton perturbation term in K-Mb solution. This can be seen an approximation solution for strong BS type perturbations on a PWB. Since the linear coherent form of a soliton and a plane wave describes the well-known interference effects in both classical wave motion theory and quantum mechanics, the evolution of $\Psi'$ can be understood well by interference effects between BS and PWB.  In physical studies or theory,  it is usual to   explain one dynamical process from other much simpler and more general description ways. Therefore, the linear interference effects, as an approximation solution, can be used to explain oscillation of strong localized perturbations on a PWB. This understanding way is similar to that MI, as an approximation solution, is used to explain amplification of weak perturbations on a PWB.  The linear superposition form is called linear  interference effect in the following text. Moreover, it should be noted the linear superposition form does not hold anymore for weak perturbation case and  intermediate cases (for which  perturbation amplitude $p$ is comparable with the background amplitude $s$). In these cases, the superposition form is a nonlinear superposition form. The nonlinear superposition form can be used to describe the  nonlinear interference process \cite{nodyzhao}, and the dominant energy of perturbations should be used to analyze the interference period \cite{zhaoling}.   In follows, we show that the interference effects can be used to explain the dynamical process of K-Mb with strong perturbations well.

 \begin{figure}[htb]
\centering
\label{fig:3}
{\includegraphics[height=50mm,width=78mm]{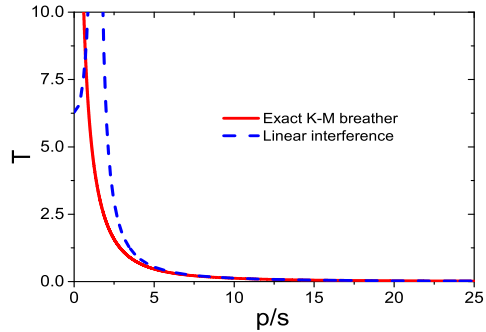}}
\caption{(color online) The oscillation period $T$ vs $p/s$ (the ratio of erturbation soliton amplitude and background amplitude). The solid red line and blue dashed line correspond to the results of K-Mb and the linear interference case respectively. It is shown that the amplitude oscillation behavior of K-M does not agree with linear interference effects for weak soliton perturbations and the  intermediate cases, but the behavior agree perfectly with the prediction of linear interference effects for strong soliton perturbations. }
\end{figure}

\section{A discussion on the mechanism of  Kuznetsov-Ma  breather}
We would like to explain the dynamical process of K-Mb from two aspects, which involve the soliton oscillation amplitude and oscillation period respectively. We calculate the maximum density value of K-Mb is $P_{max}=(s+2b)^2$, and minimum density value is $P_{min}=(s-2b)^2$. To describe the oscillation amplitude with no singularity, we define a parameter $\eta=\frac{|P_{min}-s^2|} {P_{max}-s^2}$ to describe breather's amplitude oscillation. The period of breathing is $T=\frac{2 \pi} {2 b \sqrt{b^2-s^2}}$. Then, we calculate the maximum density value of the introduced linear form $\Psi'$ is $P'_{max}=(s+p)^2$, and minimum density value is $P'_{min}=(s-p)^2$. Then characterization parameter $\eta'=\frac{|P'_{min}-s^2|} {P'_{max}-s^2}$ to describe breather's amplitude oscillation for linear interference effects between BS and PWB. The period of breathing is $T'=\frac{2 \pi} {| p^2/2-s^2|}$. Based on these defined parameters, we can compare the characters of K-Mb with the linear interference form in the two aspects, to understand the dynamical process of K-Mb. Good agreement between them means that interference effects can be used to explain the dynamical process of K-Mb, and it can be seen as a mechanism of K-Mb. This point is similar to that MI has been seen as mechanism of rogue wave and Akhmediev breather \cite{Dudley,zhaoling,Lingzhao}.

Firstly, we investigate the amplitude oscillation behavior of the two forms through plotting the defined parameter $\eta$ and $\eta'$. From the expressions of them, we can see that the ratio of BS amplitude $p$ and PWB amplitude $s$ plays essential role for determining the soliton perturbation strength is weak or strong. Therefore, we plot the defined parameter $\eta$ and $\eta'$ vs the ratio $p/s$ in Fig. 2. It is shown that the linear interference effects described line is quite different from the one described by K-Mb, for weak perturbations ($p/s<<1)$). Namely, the amplitude amplification is much larger than the value expected by interference effects. The large amplification is surely induced by MI as AB and RW.  But they agree with each other perfectly for strong perturbations ($p/s>>1$). Secondly, we show the relations between oscillation period and the ratio $p/s$ in Fig. 3. We can see that linear interference effect can still describe the period of K-Mb with strong  perturbations. But the linear interference effect fails to explain the dynamical characters for K-Mb with weak perturbations and intermediate cases (perturbation amplitude $p$  is comparable with background amplitude $s$).

\begin{figure}[htb]
\centering
\label{fig:4}
{\includegraphics[height=40mm,width=85mm]{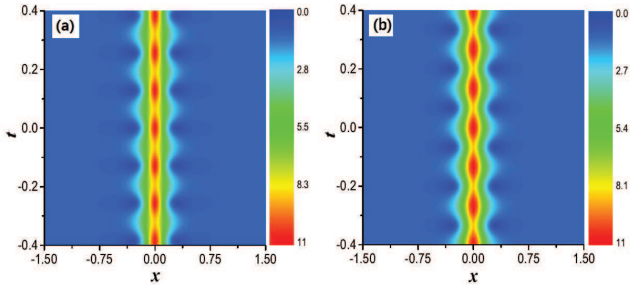}}
\caption{(color online) The evolution of K-Mb (a) and the linear interference form (b) with an identical strong perturbation strength $p/s=4 \sqrt{6}$. It is shown that their evolution processes  agree well for large perturbation soliton amplitude, which indicates the linear interference effect can be used to explain the dynamical process of K-Mb with strong perturbation cases. The plane wave background amplitude is $s=1$. }
\end{figure}

These suggest that dynamics of K-Mbs with strong perturbations can be understood by linear interference effects between BS and PWB. To support this point, we plot the evolution dynamics of K-Mb and the linear superposition form with an identical large ratio value $p/s$ in Fig. 4. The two dynamics indeed agree with each other perfectly for strong perturbation cases. Linear interference form describe the dynamics of K-Mb better with stronger BS type perturbations. For weak perturbations, the linear interference effects fails to explain the K-Mb dynamics. These cases can be understood well by linear stability analysis. The amplifications of perturbations are described well by MI mechanism. But the MI fails to explain the oscillation period of K-Mb. The oscillation period can be understood from a nonlinear interference effects between PWB and a weak perturbation. The nonlinear interference effects refer to the nonlinear superposition form of a plane wave and soliton type perturbation. In the nonlinear case, the period is determined by the evolution energy difference between the background and the perturbation's dominant energy  \cite{zhaoling}. In this way, the oscillation period can be estimated to be identical with the precise one of K-Mb solution. This is helpful to understand the recent discussion on spectral stability of rogue wave based on floquet analysis of K-Mb \cite{Cuevas}.

In fact, the MI or linear interference effect alone can not explain the dynamical process of K-Mb with perturbation amplitude $p$ is comparable with the background amplitude $s$. This can be seen from the results in Fig. 2 and Fig. 3. MI can be used to understand the amplification of amplitude, and interference effect can be used to explain the temporal  oscillation. But the breathing period does not agree with the one calculated by linear interference form anymore, since the linear superposition form does not hold  in this case. The period can be understood well with the aid of the nonlinear interference effects (the dominant energy of perturbations is also needed).   Based on the above discussions, we summarize the understanding  on dynamical process of K-Mb in Fig. 1.

The interference effects can not only explain dynamics of the K-Mb with strong BS perturbation cases, but also can be used to understand the anti-dark soliton and W-shaped soliton with nonrational form obtained in a  nonlinear Sch\"{o}dinger equation with some fourth-order effects \cite{AD}. The fourth-order effects make the soliton excitation can admit identical evolution energy with the PWB, but these cases do not exist for the typical nonlinear Sch\"{o}dinger equation and Hirota equation \cite{zhaoling,Liu1}.   Since they are reduced from a generic K-Mb, when the BS term admit identical energy with the PWB, the oscillation behavior will disappear and the breather will become an anti-dark soliton or a W-shaped soliton. Moreover, soliton excitation usually exist and RW or breather usually do not exist in MS regime. But the linear interference effect would make breather-like excitation exist in the MS regime, such as the breather-II obtained in a mixed coupled nonlinear Sch\"{o}dinger equations  \cite{defnls}. By the way, it should be noted that the interference induced breather-like excitations in MS regime can not reduce to be RW anymore, in contrast to the breathers in MI regime \cite{Kibler,Dudley,Kibler2}. This comes from that the breather-II is purely induced by the interference effects.

\section{Conclusion and discussion}
In summary,  MI and interference effect can be used to understand the dynamical process of K-Mb. MI plays dominant role for K-Mb admitting weak perturbations, and linear interference effects play dominant role for K-Mb admitting  strong perturbations. The weak and strong perturbations are clarified by the ratio $p/s<< 1$ and $p/s>>1$ respectively. For   intermediate cases, it is expected that the two mechanisms both are involved greatly. Maybe it is still need to develop some proper ways to distinct them or clarify the quantitative effects of them on the K-Mb dynamics.  The results here can be extended to explain dynamics of K-Mb and anti-dark soliton in three-wave resonant system \cite{threew}, scalar  nonlinear Schr\"{o}dinger
equation with high-order effects \cite{Hirota,SS,Wen2}, and other types nonlinear models \cite{coupledsh,1,2,6}.

An approximation form is introduced to describe the linear interference effects. The reasonability of the introduced form is  supported directly by the fact that $\Psi$ with $b$ tend to be infinity will become the form $\Psi'$.  However, it should be emphasized that the localized perturbation form is chosen to be a  $\sech$ soliton type, but it is not a generic form as the Fourier perturbation modes in linear stability analysis case.  Very recently, it was shown that many different localized perturbations could evolve to be K-Mb in a microfabricated optomechanical array \cite{Xiong}. Therefore, a more generic localized form should be introduced to explain the evolution process of strong perturbations better. Recently, many different types of strong localized perturbations on plane wave background were discussed in details, which demonstrated the evolutions process involving K-Mbs with many different periods \cite{NMI,biondini3,Biondini}.  The K-Mbs' properties and numbers can be further evaluated by calculating the inverse scattering technique eigenvalues with the initial conditions, since all eigenvalues corresponding to K-Mb and other nonlinear modes are contained in the initial conditions \cite{NA,NA2}. If  the inverse scattering eigenvalues with the initial perturbations admit continuous spectrum, the evolution of perturbation was discussed in   \cite{NMI}. If the the inverse scattering eigenvalues with the initial perturbations admit discrete  spectrum, the evolution of perturbations should correspond to the cases discussed in \cite{Zakharov,Liu}. Obviously, the inverse scattering eigenvalues for  the cases of soliton type perturbations discussed here also admit the discrete spectrum (the profile of soliton is not arbitrary for the discrete spectrum, which is different from the $\sech$-shaped ones discussed in \cite{Biondini}). But this is a qualitative understanding on nonlinear MI. A unified quantitative discussion is still needed.

\section*{Acknowledgments}
This work is supported by National Natural Science Foundation of
China (Contact No. 11775176), Shaanxi Association for Science and Technology (Contact No. 20160216),  The Key Innovative Research Team of Quantum Many-Body Theory and Quantum Control in Shaanxi Province (Grant No. 2017KCT-12), and the Major Basic Research Program of Natural Science of Shaanxi Province (Grant No. 2017ZDJC-32).

\end{document}